\documentclass[11pt,twoside]{article}  % Leave intact
\usepackage{asp2006}
\usepackage{adassconf}

\begin{document}   % Leave intact

%-----------------------------------------------------------------------
%			    Paper ID Code
%-----------------------------------------------------------------------
% Enter the proper paper identification code.  The ID code for your paper 
% is the session number associated with your presentation as published 
% in the official conference proceedings.  You can find this number by 
% locating your abstract in the printed proceedings that you received 
% at the meeting, or on-line at the conference web site.
%
% This identifier will not appear in your paper; however, it allows different
% papers in the proceedings to cross-reference each other.  Note that
% you should only have one \paperID, and it should not include a
% trailing period.
%
% EXAMPLE: \paperID{O4.1}
% EXAMPLE: \paperID{P2.7}

\paperID{O11.3}

%-----------------------------------------------------------------------
%		            Paper Title 
%-----------------------------------------------------------------------
% Enter the title of the paper.
%
% EXAMPLE: \title{A Breakthrough in Astronomical Software Development}

\title{Robust Machine Learning Applied to Terascale Astronomical Datasets}

%-----------------------------------------------------------------------
%          Short Title & Author list for page headers
%-----------------------------------------------------------------------
% Please supply the author list and the title (abbreviated if necessary) as 
% arguments to \markboth.
%
% The author last names for the page header must appear in one of 
% these formats:
%
% EXAMPLES:
%     LASTNAME
%     LASTNAME1 and LASTNAME2
%     LASTNAME1, LASTNAME2, and LASTNAME3
%     LASTNAME et al.
%
% Use the "et al." form in the case of four or more authors.
%
% If the title is too long to fit in the header, shorten it: 
%
% EXAMPLE: change
%    Rapid Development for Distributed Computing, with Implications for the Virtual Observatory
% to:
%    Rapid Development for Distributed Computing

\markboth{Ball, Brunner, and Myers}{Robust Machine Learning Applied to Terascale Astronomical Datasets}

%-----------------------------------------------------------------------
%		          Authors of Paper
%-----------------------------------------------------------------------
% Enter the authors followed by their affiliations.  The \author and
% \affil commands may appear multiple times as necessary.  List each
% author by giving the first name or initials first followed by the
% last name. Do not include street addresses and postal codes, but 
% do include the country name or abbreviation. 
%
% If the list of authors is lengthy and there are several institutional 
% affiliations, you can save space by using the \altaffilmark and \altaffiltext 
% commands in place of the \affil command.
%
% EXAMPLE: 
%      \author{Raymond Plante, Doug Roberts, 
%                  R.\ M.\ Crutcher\altaffilmark{1}}
%      \affil{National Center for Supercomputing Applications, 
%                 University of Illinois, Urbana, IL, USA}
%      \author{Tom Troland}
%      \affil{University of Kentucky, Lexington, KY, USA}
%
%      \altaffiltext{1}{Astronomy Department, UIUC}
%
% In this example, the first three authors, "Plante", "Roberts", and
% "Crutcher" are affiliated with "NCSA".  "Crutcher" has an alternate 
% affiliation with the "Astronomy Department".  The fourth author,
% "Troland", is affiliated with "University of Kentucky"

\author{Nicholas M. Ball, Robert J. Brunner, Adam D. Myers}
\affil{Department of Astronomy and National Center for Supercomputing Applications, University of Illinois at Urbana-Champaign, Urbana, IL, USA}

%-----------------------------------------------------------------------
%			 Contact Information
%-----------------------------------------------------------------------
% This information will not appear in the paper but will be used by
% the editors in case you need to be contacted concerning your
% submission.  Enter your name as the contact along with your email
% address.
% 
% EXAMPLE:  \contact{Dennis Crabtree}
%           \email{crabtree@cfht.hawaii.edu}

\contact{Nick Ball}
\email{nball@astro.uiuc.edu}

%-----------------------------------------------------------------------
%		      Author Index Specification
%-----------------------------------------------------------------------
% Specify how each author name should appear in the author index.  The 
% \paindex{ } should be used to indicate the primary author, and the
% \aindex for all other co-authors.  You MUST use the following
% syntax: 
%
% SYNTAX:  \aindex{Lastname, F.~M.}
% 
% where F is the first initial and M is the second initial (if used). Please 
% ensure that there are no extraneous spaces anywhere within the command 
% argument. This guarantees that authors that appear in multiple papers
% will appear only once in the author index. Authors must be listed in the order
% of the \paindex and \aindex commmands.
%
% EXAMPLE: \paindex{Crabtree, D.}
%          \aindex{Manset, N.}        
%          \aindex{Veillet, C.}        

\paindex{Ball, N.~M.}
\aindex{Brunner, R.~J.}
\aindex{Myers, A.~D.}

%-----------------------------------------------------------------------
%			Subject Index keywords
%-----------------------------------------------------------------------
% Enter up to 6 keywords that are relevant to the topic of your paper.  These 
% will NOT be printed as part of your paper; however, they will guide the creation 
% of the subject index for the proceedings.  Please use entries from the
% standard list where possible, which can be found in the index for the 
% ADASS XVI proceedings. Separate topics from sub-topics with an exclamation 
% point (!). 
%
% EXAMPLE:  \keywords{astronomy!radio, computing!grid, data management!workflows, 
%     instrumentation!control}


%-----------------------------------------------------------------------
%			       Abstract
%-----------------------------------------------------------------------
% Type abstract in the space below.  Consult the User Guide and Latex
% Information file for a list of supported macros (e.g. for typesetting 
% special symbols). Do not leave a blank line between \begin{abstract} 
% and the start of your text.

\begin{abstract}          % Leave intact
We present recent results from the \htmladdnormallinkfoot{Laboratory for Cosmological Data Mining}{http://lcdm.astro.uiuc.edu} at the National Center for Supercomputing Applications (NCSA) to provide robust classifications and photometric redshifts for objects in the terascale-class Sloan Digital Sky Survey (SDSS). Through a combination of machine learning in the form of decision trees, k-nearest neighbor, and genetic algorithms, the use of supercomputing resources at NCSA, and the cyberenvironment Data-to-Knowledge, we are able to provide improved classifications for over 100 million objects in the SDSS, improved photometric redshifts, and a full exploitation of the powerful k-nearest neighbor algorithm. This work is the first to apply the full power of these algorithms to contemporary terascale astronomical datasets, and the improvement over existing results is demonstrable. We discuss issues that we have encountered in dealing with data on the terascale, and possible solutions that can be implemented to deal with upcoming petascale datasets.
\end{abstract}

%-----------------------------------------------------------------------
%			      Main Body
%-----------------------------------------------------------------------
% Place the text for the main body of the paper here.  You should use
% the \section command to label the various sections; use of
% \subsection is optional.  Significant words in section titles should
% be capitalized.  Sections and subsections will be numbered
% automatically. 
%
% EXAMPLE:  \section{Introduction}
%           ...
%           \subsection{Our View of the World}
%           ...
%           \section{A New Approach}
%
% It is recommended that you look at the sample paper sample2.tex
% for examples of formatting references, footnotes, figures, equations, 
% html links, lists, and other features.  

\section{Introduction}

We summarize work carried out as part of the Laboratory for Cosmological Data Mining, a partnership between the Department of Astronomy and the National Center for Supercomputing Applications (NCSA) at the University of Illinois at Urbana-Champaign (UIUC), in collaboration with the Automated Learning Group (ALG) at NCSA, and the Illinois Genetic Algorithms Laboratory at UIUC. This combination of expertise allows us to apply the full power of machine learning to contemporary terascale astronomical datasets.

\section{Data}

The SDSS (York et al.\ 2000) is a project to map $\pi$ steradians of the Northern Galactic Cap in five broad optical photometric bands, $u$, $g$, $r$, $i$, and $z$. The Third Data Release of the survey consists of 142,705,734 unique objects, of which 528,640 have spectra. The Fifth Data Release is a superset of DR3, approximately one and a half times the size, consisting of 9.0 TB of images and 1.8 TB of flat files in FITS format. We utilize the objects with spectra as training sets and perform blind tests on subsets of the spectra. The training features are the four colors $u-g$ through $i-z$ in the four magnitude types measured by the SDSS. We provide classifications for the full DR3 sample of 143 million objects.

\section{Computing Environment}

The machine learning algorithms are implemented within the framework of the Data-to-Knowledge toolkit (D2K; Welge et al.\ 1999), developed and maintained by the ALG at NCSA. This allows the straightforward implementation of numerous $\mathrm{Java^{TM}}$ modules which automate the stages of the data mining and learning process. The machine learning algorithms available include decision tree, $k$-nearest neighbor, artificial neural network, support vector machine, unsupervised clustering, and rule association. For the terascale datasets in use here, our implementation includes enhanced versions of the standard modules which stream data of fixed type, for example single-precision floating point.

The algorithms are run on the Xeon Linux Cluster \textit{tungsten} at NCSA, as part of a peer-reviewed, nationally allocated, LRAC allocation to the LCDM project on this and other machines, renewed over multiple years. Tungsten is composed of 1280 compute nodes. Each node is a Dell PowerEdge 1750 server running Red Hat Linux with two Intel Xeon 3.2 GHz processors, 3 GB of memory, a peak double-precision performance of 6.4 Gflops, and 70 GB of scratch disk space. A further 59 TB of general scratch space is available, and the system is connected via FTP interface to the 5 PB UniTree DiskXtender mass storage system, and the nodes to each other via Myrinet. %The nodes are connected via Myricom's Myrinet cluster interconnect network.

\section{Classification}

Using decision trees, we are able (Ball et al.\ 2006) to assign the probabilities P(galaxy, star, neither-star-nor-galaxy) to each of the 143 million objects in the SDSS DR3. This enables one to, for example, either emphasize completeness (the fraction of the true number of the target object correctly identified), or efficiency (the fraction of the objects assigned a given type that are correct) in subsamples, both of which have important scientific uses.

\section{Photometric Redshifts}

\begin{figure}
\plottwo{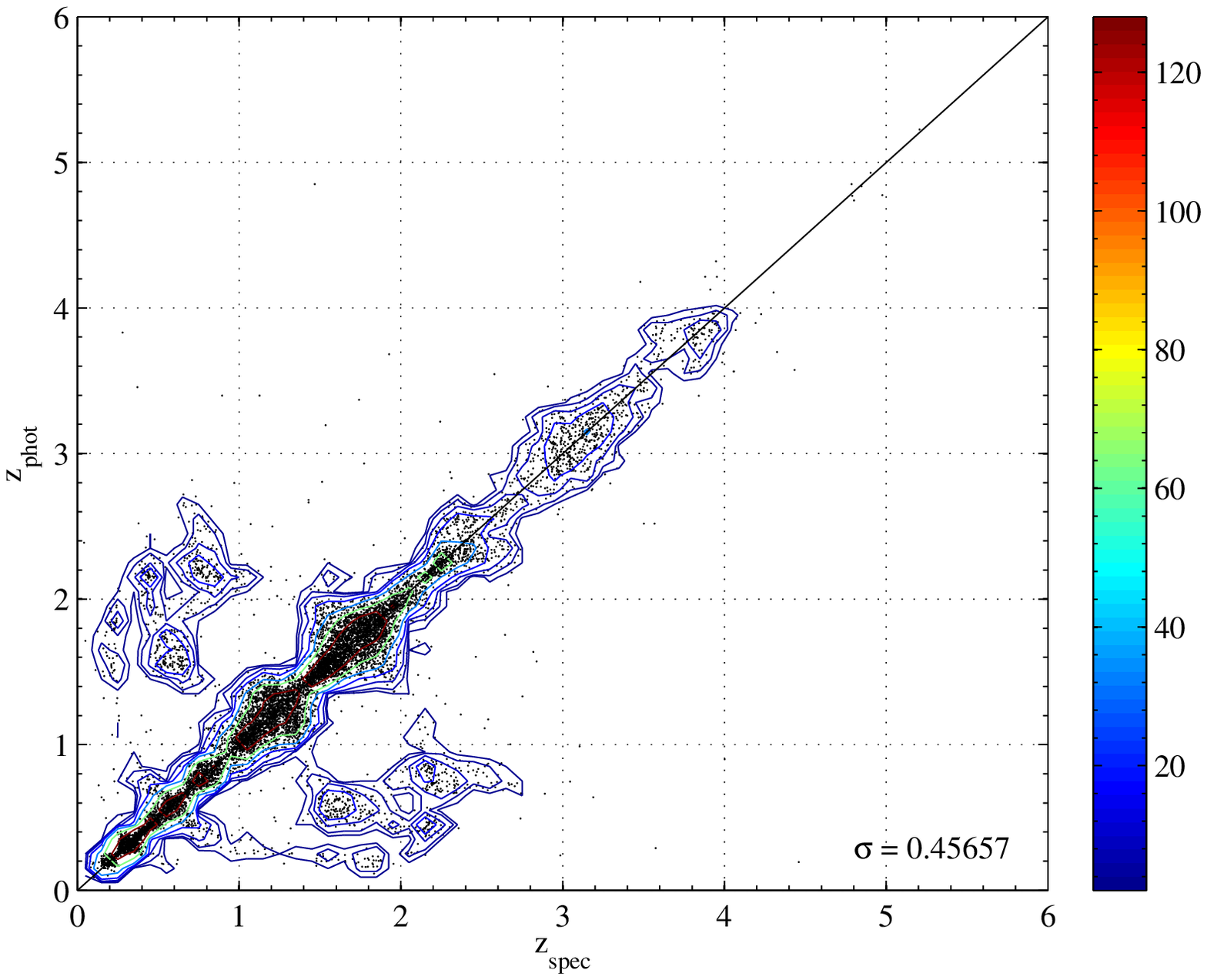}{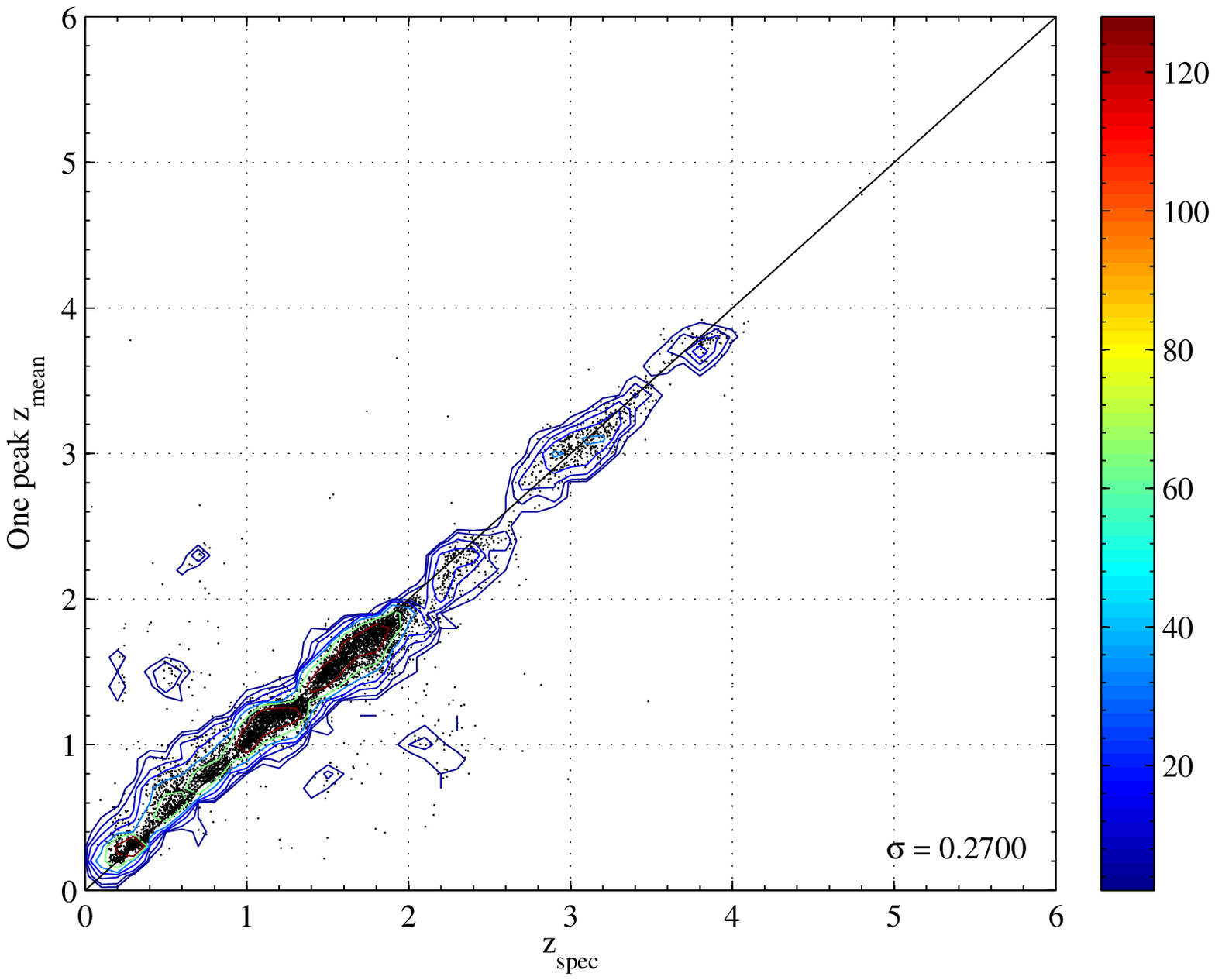}
\caption{Spectroscopic versus photometric for SDSS DR5 quasars. The left-hand panel shows a result typical of the literature until recently. The right-hand panel shows the result of using machine learning to assign probability density functions then taking the subset with a single peak in probability. \label{O11.3-fig-1}}
\end{figure}

The left-hand panel of Figure~\ref{O11.3-fig-1} shows a result typical until recently for photometric versus spectroscopic redshift for quasars, here generated as a blind test on 11,149 SDSS DR5 quasars using a single nearest-neighbor model. Most objects lie close to the ideal diagonal line, but there are regions of `catastrophic' failure, in which the photometric redshift assigned is completely incorrect. The kNN enables us to significantly reduce the instance of these catastrophics, as shown in Ball et al.\ (2007), where the RMS deviation between the photometric and spectroscopic redshifts is reduced from 0.46 as shown to 0.35.

Because every magnitude in the SDSS has an associated error, one can also perturb the testing set numerous times according to the errors on the input features, and use the resulting variation to generate full probability density functions (PDFs) in redshift. The advantage of this is that the errors on the input features are taken into account when assigning the output value. Taking the mean value from each PDF gives a similar RMS dispersion to the 0.35 result, however, for the subset of quasars which have a single PDF peak, the dispersion is further reduced to 0.27, with very few remaining outliers. This is shown in the right-hand panel of Figure~\ref{O11.3-fig-1}.

%/Users/nball/Documents/conferences/ADASS_XVII/figures/quasar_single_neighbor.eps
%/Users/nball/Documents/conferences/ADASS_XVII/figures/quasar_one_peak.eps

\section{Discussion}

Given the petascale datasets planned for the next decade, it is vitally important that contemporary data mining can be carried out successfully on this scale. In turn, this requires robust techniques on the terascale. We encountered numerous issues that were relevant to realizing this goal:

\begin{itemize}
\item Because we are using tens to hundreds of parallel nodes and streaming many GB of data, D2K must be invoked via batch script, negating the advantages of its GUI interface. The resulting lack of an integrated cyberenvironment results in batch scripts that contain many tens of settings, manually set file locations and commands, making them prone to error.
\item Job submission is inflexible, subject to fixed wallclock times and numbers of nodes, unpredictable queuing times and no recourse if a job fails due a bug or hardware problem.
\item The large datasets must be stored on the Unitree mass storage system, which is occasionally subject to outages in access or significant wait times. In combination with the queuing system for batch jobs described above, this can make new scripts time-consuming to debug.
\item There is no way in which to fully explore the huge parameter space (more than $10^{15}$ combinations of settings for decision trees) of the machine learning algorithms. Genetic algorithms were used to optimize the training features, and could be used similarly to optimize the algorithm.
\item The present lack of fainter training data forces us to extrapolate in order to classify the whole SDSS. While the data and results we obtain are well-behaved, it will always be the case in astronomy that some form of extrapolation is ultimately required. This result is simply due to the fact that photometry will always be available several orders of magnitude fainter than spectroscopy, due to the physical difficulties in obtaining spectra of faint sources. Thus while our supervised learning represents a vital proof-of-concept over a whole terascale survey, ideally it should be extended with semi-supervised or unsupervised algorithms to fully explore the regions of parameter space that lie beyond the available training spectra. It is worth noting, however, that our training features, the object colors, are largely consistent beyond the limit of the spectroscopic training set.
\item The data size is such that integrating the SQL database with D2K via JDBC is impractical, and the data must be stored as flat files. As database engines become more sophisticated, however, it could in the future become possible to offload partial or entire classification rules to a database engine. Doing so, however, would require supercomputing resources for the database engine, which results in an entirely new class of problems.
\end{itemize}

In moving to the petascale, further issues include:

\begin{itemize}
\item Conventional hardware, in the form of large clusters of multicore compute nodes, is scaleable to the petascale, however, field-programmable gate arrays (FPGAs), graphical processing units, and Cell processors may be more suited to many data mining tasks, due to their embarrassingly parallel nature. The LCDM group, in collaboration with the Innovative Systems Laboratory at NCSA, has demonstrated results on FPGAs using an SRC-6 MAPE system (Brunner, Kindratenko, \& Myers 2007), which include running a kNN algorithm, although the implementation is not trivial because the algorithm must be rewritten.
\item For many applications on the petascale, the performance becomes I/O limited. This is quantified by Bell, Gray, \& Szalay (2006), who apply Amdahl's law (Amdahl 1967) that one byte of memory and one bit per second of I/O are required for each instruction per second, to predict that a petaflop-scale system will require one million disks at a bandwidth of $100~{\mathrm{MB~s^{-1}}}$ per disk. They also state that data should be stored locally (i.e., not transferred over the internet), if the task requires less than 100,000 CPU cycles per byte of data. Many contemporary scientific applications are such that local storage is favored by over an order of magnitude.
\end{itemize}

\acknowledgments

The authors acknowledge support from NASA through grant 05-AISR05-0144.

\end{document}